\documentclass[runningheads]{llncs}
\usepackage[T1]{fontenc}
\usepackage{graphicx}
\usepackage{color}
\usepackage[unicode,colorlinks=true,allcolors=blue]{hyperref}

\usepackage[hyphenbreaks]{breakurl}
\usepackage{cite}
\usepackage{amsmath}
\usepackage{booktabs}
\usepackage{tikz}
\usetikzlibrary{calc}

\begin{document}
\title{Python Fuzzing for Trustworthy Machine Learning Frameworks}
%
%
\author{Ilya~Yegorov\inst{1,2}\orcidID{0000-0003-2158-0414} \and
Eli~Kobrin\inst{1,2}\orcidID{0000-0002-6035-0577} \and
Darya~Parygina\inst{1,2}\orcidID{0000-0002-4029-0853} \and
Alexey~Vishnyakov\inst{1}\orcidID{0000-0003-1819-220X} \and
Andrey~Fedotov\inst{1}\orcidID{0000-0002-8838-471X}}
\authorrunning{I. Yegorov et al.}
%
\institute{Ivannikov Institute for System Programming of the RAS \and
Lomonosov Moscow State University
\email{\{Yegorov\_Ilya,kobrineli,pa\_darochek,vishnya,fedotoff\}@ispras.ru}}
\maketitle              
\begin{tikzpicture}[remember picture, overlay]
\node at ($(current page.south) + (0,0.65in)$) {
\begin{minipage}{\textwidth} \footnotesize
 Yegorov I., Kobrin E., Parygina D., Vishnyakov A., Fedotov A. Python fuzzing for trustworthy machine learning frameworks. Zapiski Nauchnykh Seminarov POMI, St. Petersburg Branch of the V.A. Steklov Mathematical Institute of the Russian Academy of Sciences, Vol. 530, 2023, pp. 38–50.

Journal of Mathematical Sciences, 2024 Springer Nature Switzerland AG, Vol. 285, No. 2, October, 2024. DOI 10.1007/s10958-024-07424-2.
\end{minipage}
};
\end{tikzpicture}

\begin{abstract}
Ensuring the security and reliability of machine learning frameworks is crucial
for building trustworthy AI-based systems. Fuzzing, a popular technique in
secure software development lifecycle (SSDLC), can be used to develop secure and
robust software. Popular machine learning frameworks such as PyTorch and
TensorFlow are complex and written in multiple programming languages including
C/C++ and Python. We propose a dynamic analysis pipeline for Python projects
using the Sydr-Fuzz toolset. Our pipeline includes fuzzing, corpus minimization,
crash triaging, and coverage collection. Crash triaging and severity estimation
are important steps to ensure that the most critical vulnerabilities are
addressed promptly. Furthermore, the proposed pipeline is integrated in GitLab
CI. To identify the most vulnerable parts of the machine learning frameworks, we
analyze their potential attack surfaces and develop fuzz targets for PyTorch,
TensorFlow, and related projects such as h5py. Applying our dynamic analysis
pipeline to these targets, we were able to discover 3 new bugs and
propose fixes for them.

\keywords{Fuzzing \and Trustworthy AI \and Machine learning framework \and
TensorFlow \and PyTorch \and Python \and Artificial intelligence \and Crash
triage \and Dynamic analysis \and Secure software development lifecycle \and
SSDLC \and Computer security}
\end{abstract}

\section{Introduction}

Artificial Intelligence (AI) has been the focus of significant attention in
recent years due to its transformative potential across a wide range of
industries. However, with the increasing prevalence of AI-based systems,
security concerns have also emerged. Ensuring the reliability and safety of
these systems is paramount, and the field of secure AI has emerged as a
multi-disciplinary area of computer science dedicated to achieving this goal.

AI-based systems are typically built on top of machine learning (ML) frameworks,
such as TensorFlow~\cite{tensorflow} and PyTorch~\cite{torch}, which provide the necessary tools to implement
and train machine learning models. The growing complexity and pervasiveness of
these frameworks have made them a prime target for attackers seeking to exploit
vulnerabilities for malicious purposes. The potential consequences of a
successful attack on an AI-based system are significant, ranging from
compromising the integrity and confidentiality of sensitive data to causing
physical harm or financial loss.

Developing secure and trustworthy ML frameworks is a challenging task, given the
large and complex codebase of popular ML frameworks, which are developed in
multiple programming languages such as C/C++ and Python. Secure software development life
cycle (SSDLC) is commonly used to provide safety and code quality for open
source and commercial projects. Dynamic analysis, such as fuzzing, is often
applied in SSDLC.

Fuzzing is a method for discovering software vulnerabilities by feeding
unexpected or malformed seeds to a program. By applying fuzzing to ML
frameworks, it is possible to identify security vulnerabilities that may not be
found through traditional testing methods. However, the effectiveness of fuzzing
depends on several factors, such as the quality of the initial test cases,
the coverage of the code, and the ability to reproduce and isolate the bugs that are
discovered. Code coverage quality is also tied with fuzz targets development.
During fuzz targets development the attack surface should be considered, which
has some peculiarities for ML frameworks. Furthermore, after fuzzing, the
resulting bugs need to be triaged and prioritized to ensure that the most
critical vulnerabilities are addressed first.

This paper makes the following contributions:
\begin{itemize}
    \item We implement dynamic analysis pipeline (fuzzing, corpus minimization,
        crash triaging, and coverage collection) for Pythonic projects in
        Sydr-Fuzz~\cite{sydr-fuzz} tool.
    \item Using proposed dynamic analysis pipeline for Python, we found 3 new bugs in
        ML frameworks and related projects.
\end{itemize}

The paper is organized as follows. In Section~\ref{sec:rw} we discuss Python fuzzing tools
and approaches, ML framework specific fuzzer FreeFuzz~\cite{freefuzz}, and current fuzz targets
for TensorFlow~\cite{tensorflow} and PyTorch~\cite{torch} projects. In
Section~\ref{sec:pipeline} we describe proposed dynamic
analysis pipeline for Python projects. Section~\ref{attack-surface-analysis} is dedicated to attack surface
analysis of ML frameworks. Section~\ref{sec:evaluation} demonstrates our results of fuzzing ML
frameworks such as TensorFlow, PyTorch, and related projects.
Section~\ref{sec:conclusion} concludes
the paper.

\section{Related Work}
\label{sec:rw}

\subsection{Atheris}

Atheris~\cite{atheris} is a state-of-the-art fuzzing engine designed to guide
the testing process through code coverage analysis. Based on
libFuzzer~\cite{serebryany16}, Atheris supports fuzzing of both Python code and
native extensions for CPython. When it comes to native code, Atheris can be used
in conjunction with Address Sanitizer~\cite{serebryany12} (ASAN) or Undefined
Behavior Sanitizer (UBSAN)~\cite{sanitizers} to uncover additional bugs. A
failure criterion is triggered by an uncaught exception thrown from the Python
code or an abort caused by the sanitizers.

To collect Python coverage, Atheris provides three different bytecode
instrumentation options. The first option is to instrument the libraries that
are imported. The second option is to instrument specific individual functions.
The third option is to instrument every function currently loaded in the
interpreter. The coverage information provided by Atheris is compatible with the
popular \textit{coverage.py}~\cite{coverage} utility. Coverage reports are
generated when the fuzzer exits due to a Python exception, \texttt{sys.exit()},
or after the specified number of runs has been reached.

Fuzzing native extensions (C/C++) with Atheris is similar to fuzzing Python
code~\cite{atheris-native}. The main challenge is instrumenting the extensions
to perform effective analysis Moreover, Atheris supports custom mutators for
structure-aware fuzzing. Additionally, Atheris implements
\texttt{FuzzedDataProvider}~\cite{fuzzed-data-provider}, which facilitates the
translation of input data from simple bytes to other simple-structured forms,
such as characters, Unicode symbols, strings, integers, lists, and more. Atheris
is fully supported by OSS-Fuzz~\cite{serebryany17, oss-fuzz}.

\subsection{FreeFuzz}

FreeFuzz~\cite{freefuzz} is a novel approach to deep learning (DL) library
testing via their public APIs. This technique automates API-level fuzz testing
and provides a general and systematic approach to test DL libraries using open
source examples of API application.

Prior approaches~\cite{cradle, lemon} to testing DL libraries relied on
pre-existing models as input test cases, which could apply only a few
model-level mutations. In contrast, FreeFuzz constructs a significantly large
input space by collecting code snippets from library documentation, library
developer tests, and various DL models in the wild. The collected code is then
instrumented to trace dynamic execution information. Based on the traced data,
FreeFuzz constructs the type space, API value space, and argument value space
for the subsequent fuzzing stage.

During the main analysis stage, FreeFuzz performs mutation-based fuzzing of the
target library using several mutation rules for type and value mutations. These
rules allow to cover test cases for many API functions with a large number of
parameter values and combinations.

To resolve the test oracle problem, FreeFuzz utilizes three techniques. First,
it runs on different hardware configurations such as CPU, GPU without CuDNN, and
GPU with CuDNN to detect wrong-computation results. Second, it employs
metamorphic relations based on types precision to detect performance bugs.
Finally, it filters incorrect program terminations to distinguish crashes from
invalid seeds obtained as a result of numerous mutations.

\subsection{C++ Fuzz Targets for Tensorflow and PyTorch}

Fuzz testing is an important technique to ensure the reliability of AI
frameworks. Some fuzz targets have already been presented for the C++ side of
APIs in various frameworks, such as TensorFlow~\cite{tensorflow}. While
TensorFlow had its own simple fuzz targets, they only covered a limited amount
of code and were not very promising for fuzzing like Base64 encoding functions,
functions for path processing, functions for checking some internal structures,
etc. These fuzz targets were already added to OSS-Fuzz~\cite{oss-fuzz}. However,
TensorFlow also had some interesting fuzz targets that covered more code,
including functions for decoding and encoding different file formats such as
images and audio. Unfortunately, these targets were not added to OSS-Fuzz due to
some compilation problems. Nevertheless, we were able to build them and added
them to our repository, OSS-Sydr-Fuzz~\cite{oss-sydr-fuzz}, which is a
repository for hybrid fuzzing of open source software with our own tool,
Sydr-Fuzz~\cite{sydr-fuzz}. As a result of hybrid fuzzing these targets, we
found an error in the code responsible for decoding WAV file
format~\cite{tf-endless-loop}.

On the other hand, PyTorch~\cite{torch} did not contain any fuzz targets, and
none were added to OSS-Fuzz~\cite{oss-fuzz}. To address this issue, we wrote our
own C++ fuzz targets that covered code responsible for parsing PyTorch
intermediate representation, message deserialization, loading JIT modules, and
more. We also wrote fuzz targets for torchvision, a PyTorch subproject for
processing audio and video data~\cite{torchvision}. Our fuzz testing efforts on
PyTorch and torchvision C++ fuzz targets revealed many errors in PyTorch,
torchvision, and some of their dependencies. All Sydr-Fuzz trophies, including
errors in PyTorch, can be found in the OSS-Sydr-Fuzz trophy
list~\cite{trophies}.

\subsection{Hypothesis}

Hypothesis~\cite{hypothesis} is a Python library designed to automate code
testing. It utilizes modern property-based testing, which tests target code on a
range of input classes instead of concrete inputs.

With simple unit tests, software is typically tested on a limited number of test
cases based on the programmer's imagination. Hypothesis simplifies the process
of writing unit tests and makes them more powerful by identifying edge cases for
test code that may not be reached through human efforts.

Using Hypothesis to test code does not require significant modifications to the
unit test code. Instead, one simply needs to describe the structure of function
arguments in a way that is compatible with Hypothesis and set constraints on
them (e.g. testing code that only works with positive integer values). The test
can then be run as before, with Hypothesis generating a set of inputs based on
the given input structure and constraints. This feature enables the discovery of
corner cases that can cause the tested code to fail or behave incorrectly.

Hypothesis tests are easy to turn into fuzz targets. Along with ability to
describe the structure Hypothesis becomes a handy structure-aware fuzzing tool.

\section{Python Dynamic Analysis Pipeline}
\label{sec:pipeline}

Sydr-Fuzz~\cite{sydr-fuzz} is a powerful tool that offers a comprehensive
pipeline for dynamic analysis of Python code. This pipeline includes fuzzing,
corpus minimization, coverage collection, and crash triaging. Each component of
the pipeline is implemented as a separate command, ensuring that users can
easily access the functionalities they need.

\subsection{Fuzzing and Corpus Minimization}

Dynamic analysis is a critical component of software testing, and fuzzing is one
of the most popular methods used to carry it out. Fuzzing involves generating a
large number of inputs to a program and monitoring its behavior for unexpected
crashes or other errors. Sydr-Fuzz implements fuzzing via \texttt{sydr-fuzz run}
command and uses Atheris~\cite{atheris} fuzzing engine, which is a
coverage-guided Python fuzzing engine that supports native extensions.

Before the fuzzing session can begin, seed corpus minimization is performed to
ensure that the fuzzer initializes faster and easier with a smaller corpus. Once
the minimization is complete, the fuzzing session is launched. It continues
until the coverage stops growing for some specified period or a predetermined
number of crashes are discovered. The \texttt{exit-on-time} and \texttt{jobs}
options are responsible for managing these behaviors.

After the fuzzing session is complete, corpus minimization is the next step.
This step is crucial because the time required for all subsequent steps depends
on the size of the corpus. As such, the corpus must always be minimized to
ensure optimal performance. Additionally, minimizing the corpus may be
beneficial for reuse in subsequent launches. The \texttt{sydr-fuzz cmin} command
provides this functionality.

\subsection{Coverage Collection}

Corpus coverage is a widely accepted and fundamental metric in the context of
fuzzing. Sydr-Fuzz provides the \texttt{sydr-fuzz pycov} command, which utilizes
\textit{coverage.py}~\cite{coverage} to collect coverage information. The
\textit{pycov} command offers a range of coverage visualization formats,
including report (in the specialized coverage.py format), html, xml, json, and
lcov. While lcov is a popular choice for coverage visualization, it cannot be
used with Python egg-files due to compatibility issues with the
\textit{genhtml}~\cite{genhtml} tool. Therefore, the html format is commonly
used for Python coverage visualization.

\subsection{Crash Triaging}

Crash triaging is an essential step in the dynamic analysis pipeline. When it
comes to fuzzing, the number of crashes can be significant. Manually analyzing
each crash and identifying those that represent the same error is a laborious
and time-consuming task. To overcome this challenge, Sydr-Fuzz introduces a
\textit{casr} command that leverages the Casr~\cite{casr} toolset to automate
crash reports collection, triaging, and severity estimation. The
\texttt{sydr-fuzz casr} command has the following main stages:
\begin{enumerate}
  \item The \textit{casr-python} tool executes a fuzz target on all crashes and
    generates crash reports. In case of discovering a crash in native code, the
    \textit{casr-san} helps generate the report. The \textit{casr-python} and
    \textit{casr-san} utilize the Python crash report and sanitizer report,
    respectively. The resulting report includes valuable information such as:
    \begin{itemize}
      \item PythonReport~-- original Python crash report,
      \item CrashSeverity~-- severity estimation (exploitable, probably
        exploitable, and not exploitable),
      \item Stacktrace (Traceback),
      \item Crashline~-- source code line of crash,
      \item Source~-- source part containing the crash line,
      \item and other details.
    \end{itemize}
  \item The \textit{casr-cluster} tool performs a deduplication algorithm on the
    \textit{casr-python} reports, primarily based on the stack traces. The
    algorithm filters out standard library functions, fuzzer and sanitizer
    functions, exception handlers, and other utility functions. The crashes
    are considered identical if their stack traces are identical after
    filtering.
  \item The \textit{casr-cluster} tool clusters the deduplicated reports using
    two pseudo-metrics: TopDist and RelDist~\cite{casr-cluster}. TopDist
    measures the minimal position offset of the current frame relative to the
    topmost one, while RelDist calculates the distance between matched frames in
    two stack traces. These pseudo-metrics are used to determine the difference
    between two stack traces (\emph{similarity}). Clustering is carried out
    according to formula
    \begin{align}
      \begin{aligned}
        CLdist(CL_i, CL_j) = \underset{a \in CL_i, b \in CL_j}{max(dist(a, b))}
      \end{aligned}
    \end{align}
    Where $dist(a, b) = 1 - similarity(a, b)$, $CL_i$ and $CL_j$~-- different
    classes.
\end{enumerate}

The crash triaging process generates a summary of clusters, with each cluster
summary detailing the number of crashes it contains and a brief description of
each crash, including error information (Python exception or sanitizer error)
and crash line. Furthermore, the reports for each of the triaged crashes are
also provided.

This cluster summary provides an overview of the identified crashes, allowing
developers to focus on the most critical and frequent ones. The description of
each crash with error information and crash line helps in understanding the root
cause of the issue and the code segment responsible for the crash. Moreover, the
individual reports for each triaged crash can provide further insights into the
specific problem encountered and aid in resolving the issue.

Overall, the summary of clusters and individual reports generated by Casr can be
a valuable resource in the debugging process and help developers save time by
prioritizing the most severe and frequent crashes.

\subsection{Continuous Integration}

Continuous fuzzing integration utilizing the Sydr-Fuzz~\cite{sydr-fuzz}
framework can be automated using GitLab CI, as demonstrated by the
OSS-Sydr-Fuzz~\cite{oss-sydr-fuzz} project. It requires the following prepared
files:
\begin{itemize}
  \item a Dockerfile to build your project and fuzz targets,
  \item fuzzing configuration TOML-files,
  \item target seed corpus and dictionaries.
\end{itemize}

To prepare fuzz targets, developers should build the project with the address
sanitizer to detect crashes in native extensions and import the Atheris library
for Python fuzzing. The CI artifacts include an archive of triaged crashes and
their corresponding Casr~\cite{casr} reports, the corpus, collected coverage,
and log files for each step of the pipeline.

The fuzzing process is launched when an external CI trigger event occurs. The
Docker container is built (utilizing all cores) and used for all project fuzz
targets, and each fuzz target is launched through all dynamic pipeline steps.
All fuzz targets are analyzed in parallel. Each dynamic analysis step utilizes
at most 4 cores for a single fuzz target. The fuzzing job output
includes resulting corpus, coverage information, Casr reports, and error
triggering seeds.

By automating the fuzzing process, developers can more efficiently identify and
address potential bugs and vulnerabilities, improving the overall quality and
security of the software. The use of GitLab CI and the proposed toolset
simplifies the process and reduces the time and resources required for effective
fuzzing.

\section{Attack Surface Analysis}
\label{attack-surface-analysis}

To effectively fuzz the code of AI frameworks, it is important to identify the
potential attack surface and narrow down our focus to the parts we are
interested in.

Firstly, we can divide the implementation of the frameworks into two main parts:
the kernel that contains the core functionality, and the ecosystem which
includes both the framework's own extensions and third-party libraries that are
based on the framework.

Secondly, while the major functionality of AI frameworks is implemented using
low-level languages such as C/C++, they also provide programming interfaces for
Python. These interfaces are generated using code generation, extensions, and
other binder libraries.

Thirdly, AI frameworks rely on several third-party software modules for
implementing their kernel functionality. These dependencies may include
libraries for data serialization and deserialization, data processing, and more.
Errors in such dependencies can be critical and therefore, we need to consider
them as well.

To conduct our analysis, we focus on identifying errors in the Python API using
Sydr-Fuzz~\cite{sydr-fuzz} with Atheris~\cite{atheris} engine. We create Python
fuzz targets that cover the kernel interfaces, which internally interact with
the kernel program interfaces written in C/C++, through Python bindings. This
approach enables us to test both the C/C++ and Python code simultaneously.
Additionally, we also prioritize fuzzing the Python API of external libraries
from the AI ecosystem.

\section{Evaluation}
\label{sec:evaluation}

We applied our Python dynamic analysis pipeline to two of the most widely used
machine learning frameworks, TensorFlow~\cite{tensorflow} and
PyTorch~\cite{torch}. Additionally, we explored the TensorFlow ecosystem and
found that the h5py~\cite{h5py} tool was well-suited for fuzzing with
Sydr-Fuzz~\cite{sydr-fuzz}. By utilizing our dynamic analysis pipeline, we were
able to comprehensively analyze these frameworks and identify potential
vulnerabilities, improving the security and reliability of these widely used
tools.

\subsection{Experimental Setup}

Our experiments were conducted on a machine with two AMD EPYC 7542 32-Core
processors and 512GB RAM, running Ubuntu 20.04 and Python 3.8. To ensure
consistent and reproducible results, all analysis stages were launched within
Ubuntu-based Docker containers with the target projects pre-built and
installed~\cite{oss-sydr-fuzz}. The Docker build stage was optimized to utilize
all available cores, while each fuzz target analysis was restricted to using at
most 4 cores. In order to balance the need for comprehensive testing with
practical constraints, we stopped fuzzing when there was no new coverage for one
hour or when the total fuzzing time exceeded 24 hours.

\subsection{PyTorch}

We conducted an analysis of PyTorch~1.13~\cite{torch}, an open-source machine
learning framework, through its Python API. After examining the kernel with core
functionality, we determined that the PyTorch JIT module was the most suitable
for writing fuzz targets, specifically focusing on the \texttt{torch.jit.load}
function, which loads a \texttt{ScriptModule} or \texttt{ScriptFunction} that
has been previously saved with \texttt{torch.jit.save}.

To perform our fuzzing process, we built the PyTorch project from source and
linked it to the libFuzzer library, adding necessary instrumentation for ASAN
support (\texttt{-fsanitize=fuzzer-no-link,address}). We then installed the
produced \texttt{.whl} package via \texttt{pip}.

Our fuzz target for the Python API function~\cite{torch-targets} involved calling
\texttt{torch.jit.load} with a file containing input bytes as a parameter.
Utilizing four cores, we discovered two crashes~\cite{torch-crash-1,
torch-crash-2} during the fuzzing process, both of which were related to null
pointer dereference and SEGV due to read memory access in a third-party PyTorch
dependency called \textit{flatbuffers}. Subsequently, we were able to find these
same two crashes by fuzzing the C++ PyTorch API with Sydr-Fuzz~\cite{sydr-fuzz}
via AFL++~\cite{fioraldi20} fuzzer. We submitted a pull request to PyTorch in
order to address these issues~\cite{torch-pr}.

\subsection{h5py}

The HDF5~\cite{hdf5} binary data format is a crucial part of the AI ecosystem,
and h5py~\cite{h5py}, a Pythonic interface to it, is widely used in
TensorFlow~\cite{tensorflow} module Keras~\cite{keras} for saving and loading
models in HDF5 format.

To prepare for the dynamic analysis pipeline, we built and installed the
h5py~3.8.0
project in a specific way. First, we built the HDF5 project, the C core of h5py,
from source and linked it to libFuzzer library. We also added necessary
instrumentation for ASAN support (\texttt{-fsanitize=fuzzer-no-link,address}).
Second, we built h5py itself in the same way using the preliminary built HDF5
core module. Finally, we installed the produced \texttt{.whl} package using
\texttt{pip}.

Since the library's purpose is to manage HDF5 files, we chose the \texttt{File}
object constructor as the fuzz target~\cite{h5py-targets}. The fuzz target contained
\texttt{h5py.File} object initialization that took a file with input data in
HDF5 format as a parameter. During the fuzzing process on 4 cores, we discovered
one crash that led to an out-of-bounds access on read~\cite{hdf5-crash}. It is
noteworthy that this crash was also found by fuzzing the C-project HDF5 itself
through \texttt{H5Fopen/H5Dopen2} functions. We submitted a pull request to HDF5
in order to address this issue~\cite{hdf5-pr}.

\subsection{TensorFlow}

We applied the Sydr-Fuzz Python dynamic analysis pipeline to
TensorFlow~2.11~\cite{tensorflow}, one of the most popular open-source machine
learning platforms. After building the project from source and linking it to the
libFuzzer library with added ASAN support, we extensively researched the
TensorFlow Python API and wrote fuzz targets for modules such as \texttt{io},
\texttt{audio}, \texttt{keras}, \texttt{image}, and
\texttt{strings}~\cite{tensorflow-targets}. Despite
our efforts, no crashes were found during the fuzzing process.

\subsection{Crash Triaging with Casr}

Casr~\cite{casr} was used to simplify crash analysis. Initially, Python reports
were generated for the crashes that were found. The reports were then
deduplicated and clustered, and severity estimation was performed. Casr was run
on 4 cores.

Table~\ref{tbl:casr} shows that Casr was effective in reducing the number of
crashes. For PyTorch, the tool reduced the number of crashes from 345 to 2 that
were significantly different, while for h5py, it reduced the number of crashes
from 190 to just 1. Generating the reports was the most time-consuming part of
the process, as each crashing seed required a fuzz target run, which had a
long Atheris initialization stage. However, report deduplication and clustering
were quick, taking only a couple of seconds.

\begin{table}
    \caption{Crash Triaging with Casr (4 cores).}\label{tbl:casr}
\centering
\begin{tabular}[htbp]{l r r r}
  \toprule
  \textbf{Fuzz Target}              & PyTorch (load) & h5py (file) \\
  \midrule
  \textbf{Number of Crashes}        & 345            & 190 \\
  \textbf{Deduplicated Reports}     & 2              & 1   \\
  \textbf{Number of Clusters}       & 2              & -   \\
  \midrule
  \textbf{Reports Collection (sec)} & 3767           & 601 \\
  \textbf{Deduplication (sec)}      & 2              & 1   \\
  \textbf{Clustering (sec)}         & 1             & -   \\
  \bottomrule
\end{tabular}
\end{table}


\section{Conclusion}
\label{sec:conclusion}

In conclusion, our analysis of different approaches to testing AI frameworks led
us to propose our own approach utilizing Sydr-Fuzz~\cite{sydr-fuzz}, which
includes a sequential pipeline of coverage-guided fuzzing with
Atheris~\cite{atheris}, corpus minimization, coverage collection, and crash
triaging with Casr~\cite{casr} that includes crash deduplication, clustering,
and preparing human-friendly convenient  reports. This pipeline was automated
with GitLab CI.

We chose to focus on fuzzing Python APIs of AI frameworks and tools from their
ecosystem, building C/C++ parts of libraries with libFuzzer~\cite{serebryany16}
and sanitizers instrumentation to provide full-fledged fuzzing. We settled on
fuzzing \texttt{load} function from PyTorch JIT module, file opening module from
h5py, and some functions for parsing different formats from TensorFlow. Through
this approach, we were able to find crashes in C/C++ parts of PyTorch and h5py
through their Python APIs, ultimately deduplicating and clustering them into a
manageable number of crashes: 2 crashes in
PyTorch~\cite{torch-crash-1,torch-crash-2} and 1 crash in
h5py~\cite{hdf5-crash}. All the found crashes were patched by corresponding pull requests~\cite{torch-pr,hdf5-pr}.

\bibliographystyle{splncs04}
\bibliography{bibliography}

%
%
%
%
\end{document}